\documentclass[12pt]{article}
\usepackage{amsmath, amssymb, graphicx, subcaption, natbib}
\usepackage{url}
\usepackage{xcolor}
\usepackage{setspace}
\usepackage{natbib}
\usepackage{enumerate}
\usepackage[top=2cm,left=2cm, right=2cm, footskip=2cm]{geometry}

\newcommand{\edit}[1]{\textcolor{black}{#1}}
\usepackage{hyperref}
\usepackage{xr}

\title{A likelihood-based framework for {simultaneously} learning {both} noise and growth dynamics using biologically-informed neural networks}
\author{Rebecca M. Crossley$^{1}$ and Ruth E. Baker$^{1}$}
\date{}
\begin{document}

\maketitle
\begin{center}
{$^{1}$\emph{Mathematical Institute, University of Oxford, United Kingdom OX2 6GG }}
\end{center}
\begin{abstract}
In recent years, neural ordinary differential equation frameworks such as Biologically-Informed Neural Networks (BINNs) have shown promise for learning mechanistic laws from sparse data. 
However, most {existing approaches implicitly} assume homoscedastic Gaussian noise, {and therefore do not account for potentially} meaningful structure in {biological} variability. 
Here, we present an extension to the existing BINNs framework that includes a learnable noise model, allowing discovery of {the noise model} directly from data. 
Using population growth as an example, we demonstrate that the framework accurately recovers the underlying noise structure {and improves} predictions of the underlying growth laws compared to existing {approaches}.
As such, this work establishes a general likelihood-based framework for jointly learning dynamics and heteroscedastic noise within mechanistic neural network approaches.
\end{abstract}

\section{Introduction}
The rapid growth in data availability over recent years has driven increasing interest in hybrid modelling approaches that combine mechanistic insight with the flexibility of machine learning. 
{
Despite this, learning mathematical equations from real-world data remains challenging: observations are often noisy, sparse, and only partially informative of the underlying dynamics. 
This is particularly true in biological systems, where variability and uncertainty are intrinsic features of the data.}


{C}onsider a simple ordinary differential equation (ODE) model that describes population growth,
\begin{equation}
\label{equation:PINNs}
\frac{\mathrm{d}{u}}{\mathrm{d}{t}}= u\, g(u; \boldsymbol{p}),
\end{equation}
where $u(t)$ denotes the population density, {and $g(u; \boldsymbol{p})$ represents a density-dependent growth law with unknown parameters} $\boldsymbol{p}$~\cite{murray_mathematical_1993, baker2025modeling}. 
In classical modelling, the form of $g(u; \boldsymbol{p})$ is specified {\emph{a priori}} and parameters are {subsequently} estimated from data.
However, rather than specifying all components of the model equations in advance, unknown relationships {or parameters} in the system can instead be represented using universal function approximators and learned directly from data.

In neural differential equation frameworks, neural networks are used as these approximators~\cite{Raissi2019physics, Pang2020, lagergren2020biologically, nardini2024forecasting}{.}
{Physics-Informed Neural Networks (PINNs)} are {one} prominent example of this approach {that embed} known governing equations directly into the loss function to enable both data-driven prediction of system dynamics and inverse parameter inference from sparse or noisy data~\cite{Raissi2019physics, Pang2020}.
The success of PINNs has inspired a broader family of domain-aware learning frameworks that incorporate prior knowledge through differential operators, penalty functions, or architectural constraints.
{Biologically-Informed Neural Networks (BINNs)}~\cite{lagergren2020biologically} extends the PINN framework to settings where mechanistic structure is only partially known and data are limited, as is often the case in biology.
BINNs address these challenges by incorporating soft, biologically motivated constraints such as conservation laws, rate equations, or population-level interactions directly into the loss function, yielding models that are both flexible and interpretable even in sparse regions of data.
{For example,} one neural network may be used to approximate the solution, $u_\theta(t)$, whilst others represent unknown components of the governing equations, such as $g_\phi(u)$. 
These quantities are learned simultaneously by minimising a shared loss function that penalises disagreement with both the observed data and the governing equations. 
In this way, the model learns dynamics that are consistent with both the observations and the prescribed mechanistic structure, without requiring that the governing equations are solved explicitly.


A range of alternative methods have been proposed for discovering dynamical equations from data, including sparse identification of nonlinear dynamics, symbolic regression, Gaussian processes, and variational inference frameworks~\cite{baker2025modeling, archambeau2007gaussian,  udrescu2020ai, brunton2016discovering}. 
While these methods have demonstrated success in specific settings, many rely on restrictive assumptions, such as predefined function libraries of terms or dense, low-noise observations.
In contrast, BINNs avoid explicit model specification by learning functional relationships directly{, making them} well suited to sparse and noisy biological data.

To date, most applications of BINNs have focused on identifying mechanisms in complex {biological} systems {that exhibit both spatial and temporal variation}, such as {those modelled using} reaction--diffusion equations or stochastic agent-based simulations{~\cite{lagergren2020biologically, nardini2024forecasting, lavery2026physics}. BINNs have also been adapted to learn underlying biological mechanisms from} large-scale omics data~\cite{ ferraro2023movida, hartman2023interpreting, hossain2024biologically, selby2025beyond}. 
By contrast, BINNs' use for discovering and interpreting ODE models remains comparatively under explored, despite the centrality of ODEs in population modelling and systems biology \cite{daneker2023systems, su2021deep, yazdani2020systems}. 

A key limitation of most existing neural ODE and BINN approaches is their treatment of observation noise. 
Biological data are inherently noisy, and {the magnitude of this variability often} depends on the state of the system, leading to heteroscedastic noise structures.
Despite this, most existing neural ODE based approaches implicitly assume fixed noise models, such as constant-variance Gaussian noise through the use of mean squared error losses. 
{As a result, variability is typically treated as a nuisance to be minimised, rather than as a potentially informative component of the underlying dynamics.}
While heteroscedastic noise modelling is well established in statistical learning~\cite{tran2024robust, immer2023effective, rutemiller1968estimation}, its integration into mechanistic neural differential equation frameworks remains limited\edit{~\cite{kendall2017uncertainties, pilar2024physics}}.


In this work, we address this limitation by introducing the {negative log-likelihood trained BINN (NLL--BINN)} framework which embeds a parametric noise model directly within a likelihood-based formulation of a BINN. 
This allows both the governing dynamics and the structure of observation noise to be inferred simultaneously from data, without requiring predefined libraries or fully Bayesian inference schemes. 
This distinction is important: rather than treating uncertainty purely as a statistical artefact, the proposed approach allows the noise structure itself to be interpreted as a function of the population density, thereby linking variability to underlying biological mechanisms. 
As such, the method occupies a complementary position to probabilistic neural ODE and Bayesian approaches, trading full posterior inference for a computationally efficient and interpretable representation of heteroscedastic noise within a mechanistic learning framework.

Specifically, we introduce a power-law noise model that captures additive, multiplicative, and intermediate heteroscedastic noise {(\textit{i.e.}, noise whose magnitude scales sub-linearly with the population density)} within a unified formulation, and thus is capable of recapitulating the results of existing BINNs, whilst also extending their use to consider new meaningful noise models. 
To demonstrate the utility of NLL--BINNs, we apply it to {both synthetic datasets and real experimental data, including coral reef re-growth observations from the Great Barrier Reef, where variability cannot be estimated from replicated measurements.}
We focus on classic sigmoidal models including the {l}ogistic, Gompertz, and Richards’ equations, which are specifically chosen because they produce similar population-level dynamics despite having different underlying crowding laws \cite{gutenkunst2007universally}.
Using synthetic datasets with known ground truth, we demonstrate that the proposed approach can accurately recover both the underlying growth laws and the associated noise scaling.
Importantly, {our method demonstrates that} incorporating a learnable noise model enables not only improved uncertainty quantification, but also deeper insight into the mechanisms governing variability in biological systems. 
By linking noise structure to population size, the NLL--BINN framework moves beyond treating uncertainty as a by-product of modelling, but instead repositions it as an integral component of mechanistic inference.




\section{The NLL--BINN framework}
Building on the general BINN formulation introduced above, we now introduce a likelihood-based extension in which {the structure of} the observation noise is learned jointly with the system dynamics, rather than being prescribed \emph{a priori}. 
{This extension is particularly important in biological systems, where observational data are inherently stochastic due to measurement error, process noise, and other sources.}

{A common assumption in mechanistic modelling is that observations ${{u^\textrm{o}_i}}$ at time $t_i$ with $i\geq0$ contain additive Gaussian noise with constant variance, given by
\begin{equation}
u_i^{\mathrm{o}} = u(t_i) + \eta_i,
\end{equation}
where
\[
\eta_i \sim \mathcal{N}(0,\sigma^2).
\]
Under this assumption, maximum likelihood estimation corresponds to minimising a mean squared error (MSE) loss between the observations and the model predictions.}
{However}, biological data are often heteroscedastic, meaning that the magnitude of the noise {varies between observations}. 
To account for this, we instead model the {observed data using}
\begin{equation}
u_i^{\mathrm{o}} = u(t_i) + \epsilon\,\sigma(u(t_i)),
\label{eq:data_struc}
\end{equation}
where $\epsilon \sim \mathcal{N}(0,1)$ {and $\sigma(u(t_i))$ controls how the magnitude of the observational variability depends on the population density.} 

To capture {density dependence} in a simple and interpretable manner, we assume a power-law form for the {magnitude of the} noise,
\begin{equation}
\label{eq:power_noise}
\sigma(u) = \sigma_0 |u|^{\alpha},
\end{equation}
where $\sigma_0 > 0$ is a scale parameter and $\alpha$ controls how the noise magnitude depends on the population density (although alternative {functional forms} could be considered where appropriate). 
When $\alpha=0$, Equation~\eqref{eq:power_noise} describes additive {Gaussian} noise, corresponding to constant measurement variance; whereas when $\alpha=1$, the model corresponds to multiplicative noise, where the standard deviation scales linearly with the mean. 

To account for the data structure in Equation~\eqref{eq:data_struc}, the NLL--BINN framework learns both the true underlying solution, and the noise model in the data simultaneously.
{In particular, the noise parameters} $\alpha$ and $\sigma_0$ are {treated as learnable quantities}, enabling discovery of the noise scaling law directly from the {observations}.

{The smooth underlying solution, {$u(t)$}, is approximated using a neural network{, $u_\theta(t)$,} while a second network{, $g_\phi(u)$,} is used to represent the unknown growth law, {$g(u)$}, in Equation~\eqref{equation:PINNs}.
Both networks are trained jointly, with automatic differentiation used to enforce the governing equation as a soft constraint during optimisation.}

During training, the NLL--BINN simultaneously tries to fit the observed data to the {learned solution $u_\theta(t)$,} the {governing equation} (Equation~\eqref{equation:PINNs}),
{and} {ensures} that any other biological knowledge, such as non-negativity of population densities, is also satisfied. 
{A loss} function {of the following form} is used to do this: 
\begin{equation}  
\label{eq:Loss}
\mathcal{L}_{\mathrm{{t}otal}} = w_{\mathrm{data}}\mathcal{L}_{\mathrm{data}} + w_{\mathrm{ODE}}\mathcal{L}_{\mathrm{ODE}} + w_{\mathrm{bio}}\mathcal{L}_{\mathrm{bio}},
\end{equation}
where $\mathcal{L}_{\text{data}}$ {penalises differences between the {observations} and the learned} solution, $u_{\theta}(t)$; $\mathcal{L}_{\text{ODE}}$ penalises any discrepancies between the left and right hand side{s} of Equation~\eqref{equation:PINNs} {(when evaluated using $u_\theta(t)$ and $g_\phi(u)$)}; and $\mathcal{L}_{\text{bio}}$ enforces soft biological constraints such as positivity {of densities}.
The {weights} $w_{\mathrm{data}}, w_{\mathrm{ODE}}, w_\mathrm{bio}\ge0$ control the {relative} contribution{s} {of each term} in the total loss function (Equation~\eqref{eq:Loss}).
Taken together, the loss combines fidelity to data, agreement with the governing ODE, and avoidance of physically implausible outcomes in both the solutions and the learned dynamics. 
In this work, {we adopt} equal weighting, $w_{\text{data}} = w_{\text{ODE}} = w_{\text{bio}} = 1$, which provide{d} stable and consistent training across all experiments. 
We now describe each component of the total loss in more detail.

\paragraph{The data loss.}
The BINN is trained to maximise the likelihood of the observed data under the noise model prescribed in Equation~\eqref{eq:power_noise} whilst jointly satisfying the other constraints. 
Assuming that {$\epsilon\sim \mathcal{N}(0,1)$ and that {the measurement noise has} density-dependent variance defined by Equation~\eqref{eq:power_noise}, then} the corresponding negative log-likelihood (NLL) loss for the data {(as described in Equation~\eqref{eq:data_struc}) term in Equation~\eqref{eq:Loss}} is given by
\begin{align}
\mathcal{L}_{\mathrm{data}}
&= -\frac{1}{N}\sum_{i=1}^{N} \log \left[
\frac{1}{\sqrt{2\pi\,(\sigma(u_\theta(t_i)))^2}}
\exp\left(
-\frac{\big({{u^\textrm{o}_i}} - u_\theta(t_i)\big)^2}{2(\sigma(u_\theta(t_i)))^2}
\right)
\right] \nonumber \\
&= \frac{1}{N}\sum_{i=1}^{N} 
\left[
\frac{\big({{u^\textrm{o}_i}} - u_\theta(t_i)\big)^2}{2(\sigma(u_\theta(t_i)))^2}
+ \frac{1}{2}\log\!\big(2\pi(\sigma(u_\theta(t_i)))^2\big)
\right].
\label{eq:nll_loss}
\end{align}
In practice, the variance is evaluated at the observed data ${{u^\textrm{o}_i}}$ rather than the latent state $u_\theta(t_i)$, which provides a stable and fully observable approximation during training~\cite{lagergren2020biologically}. 
This formulation allows the BINN to account for both additive and multiplicative noise within a unified probabilistic framework, while preserving the ability to compute calibrated uncertainty bands during inference.
{Importantly, this framework is not restricted to Gaussian noise, and alternative distributions could be incorporated by specifying the corresponding likelihood function.}

\paragraph{The ODE loss.}
To {encourage consistency with} the governing {equation, in this case Equation~\eqref{equation:PINNs},} is satisfied across the entire temporal domain, and not just at the measured data points, we randomly sample $N_{\mathrm{ODE}}$ time points $t^{\mathrm{ODE}}_i \in [0,t_{\mathrm{end}}]$, where $t_\mathrm{end}$ is the final time point of interest. 
At each sampled point, the ODE residual is evaluated by substituting the MLP approximations $u_\theta$ and $g_{\phi}$ into Equation~\eqref{equation:PINNs}. 
The ODE constraint loss $\mathcal{L}_{\mathrm{ODE}}$ is therefore defined by
\begin{align}
\label{eq:ODE_loss}
\mathcal{L}_{\mathrm{ODE}}&= \frac{1}{N_{\mathrm{ODE}}} \sum_{i=1}^{N_{\mathrm{ODE}}} 
\Bigg( 
    \frac{{\mathrm{d}} }{{\mathrm{d}} t}u_{\theta}\left(t^{\mathrm{ODE}}_i\right)- u_{\theta}\left(t^{\mathrm{ODE}}_i\right) \,
      g_{\phi}\left(u_{\theta}\left(t^{\mathrm{ODE}}_i\right)\right)\Bigg).
\end{align}

\paragraph{The biological loss.}
To ensure that the learned solutions and operators are biologically realistic, we incorporate a biological loss term $\mathcal{L}_{\text{bio}}$. 
This term could include restrictions on the functions, such as ensuring that $g_\phi$ is \edit{monotonic or} bounded but{,} for simplicity, {in this work} we constrain the values of the learned quantity, $u_\theta(t)$,  to ensure non-negativity.
As such, 
\begin{equation*}
\mathcal{L}_{\text{bio}} =
\frac{1}{N_{\text{bio}}}
\sum_{i=1}^{N_{\text{bio}}}
\left(\max\left(0,\,-u_\theta(t_i^{\mathrm{bio}})\right)\right)^2
,
\end{equation*}
where the sum is taken over $N_{\text{bio}}$ randomly sampled points $t_i^\text{bio}\in [0,t_{\mathrm{end}}]$. 
This penalises the network only when the predicted quantity, $u_\theta$, {become{s} negative} and falls outside of the biologically meaningful range.

\vspace{.5cm}
Importantly, given the structure of the data loss term, the total loss function in Equation~\eqref{eq:Loss} can be interpreted as a penalised likelihood {function}. 
The data term corresponds to the NLL under a Gaussian noise model with density-dependent variance, while the ODE and biological terms act as regularisation penalties that enforce mechanistic consistency and biological plausibility. 
From this perspective, the NLL–BINN framework performs maximum likelihood estimation subject to soft constraints imposed by the governing equations and prior knowledge. This interpretation provides a useful link between mechanistic neural differential equation models and classical approaches in statistical inference and inverse problems.

\subsection{Training procedure and hyperparameters}
In the implementation used here, both the trajectory network, $u_{\theta}$, and the growth law network, $g_{\phi}$, are MLPs with three hidden layers, each containing sixty-four nodes.
They all use \texttt{tanh} activations by default. 
Training is performed with an Adam optimiser on the CPU. 

Before training, the data are randomly partitioned into training (60\%), validation (20\%), and held-out test (20\%) subsets using a fixed random permutation.
During optimisation, only the training dataset is used as $u^\textrm{o}$, which is subdivided into randomly selected mini-batches of size 64 which are reshuffled each epoch. 
The best neural network configuration is selected and saved using the smallest validation loss (calculated by executing the network configuration at each epoch on the validation dataset).

For all \edit{synthetic} experiments, we use ten data replicates, and the data split into training, validation and test sets are fixed independently within each replicate. 
The NLL--BINN is trained \edit{with the same training/validation data split} for 4000 epochs for an ensemble of {ten} independent BINNs each with a different random seed to account for varying possible initialisations.
\edit{For real data in Section~\ref{sec:CR}, we have only one data replicate. Again, the data is split into training, validation and test sets prior to training, and an ensemble is generated using ten independent BINN initialisations.}
Summary statistics such as the ensemble mean and standard deviation are reported for the predicted trajectories, the growth law, and the learned noise level to improve reliability in the results. 

By default, the parameters used in the training implementation are \edit{chosen based on previous studies and tested to ensure robustness in this specific problem setting~\cite{crossley2026reliable}. The choices included} a learning rate of \(10^{-3}\) \edit{and} the number of {auxiliary (ODE collocation) time points per iteration, {$N_{\mathrm{ODE}}$, taken as $256$}.
{Notably, the auxiliary time points are {randomly sampled in $[0, t_\mathrm{end}]$ and therefore are} distinct from the observed data (where Equation~\eqref{eq:nll_loss} is evaluated) and are used in Equation~\eqref{eq:ODE_loss} to ensure that the governing equation is satisfied across the entire time domain.}
Code and training scripts are available at: \url{github.com/beckycrossley/Learning-noise-growth-BINNs}.


\section{Mathematical models}\label{sec:models}
{In order to evaluate the ability of the NLL--BINN framework to recover underlying dynamics and noise structure}, this work focuses on three classical population growth models that can all be expressed in the general form of Equation~\eqref{equation:PINNs}{: the logistic, Gompertz and Richards' models}.
{These models are used solely to generate synthetic datasets with known ground truth, allowing quantitative assessment of model recovery and uncertainty calibration.}
One key feature of all three of these models is that the crowding functions each differ in form, yet generate qualitatively similar sigmoidal growth dynamics \cite{gutenkunst2007universally} that are initially difficult to tell apart \citep{simpson2022parameter}.
{Importantly though, the underlying functional forms of these models are not provided to the NLL--BINN at any point during training; instead, the framework infers these dynamics directly from the noisy data.}
We introduce these three models below\edit{, and data generated from these models can be seen in Figure~\ref{fig:intro} (middle row).}

\paragraph{The {l}ogistic model.}
In the {l}ogistic growth law \cite{murray_mathematical_1993}, the crowding function is linear in the population density:
\begin{equation}
    g(u) = r \left(1 - \frac{u}{K}\right), \label{eq:log_g}
\end{equation}
where $r>0$ represents the intrinsic growth rate, and $K>0$ represents the carrying capacity of the population, such that $g(K)=0$ and $g(0)=r.$
For an initial population given by $u(0) = u_0$, then the explicit solution of Equation~\eqref{equation:PINNs} subject to the {l}ogistic growth law in Equation~\eqref{eq:log_g} is given by
\begin{equation}
    u(t) = \dfrac{K}{1 + \left(\dfrac{K}{u_0} - 1\right)e^{-rt}}. \label{eq:log_sol}
\end{equation}

\paragraph{The Gompertz model.}
The Gompertz model~\cite{winsor1932gompertz} instead employs a logarithmic crowding function given by
\begin{equation}
    g(u) = r \ln\!\left(\dfrac{K}{u}\right), \label{eq:Gompertz_g}
\end{equation}
where, again, $r>0$ is the intrinsic growth rate, and $K>0$ is the carrying capacity of the population.
In this case, $g(K)=0$ but $g(u)\to\infty$ as $u\to0^{+}$.
For initial condition $u(0)=u_0$, the explicit solution is
\begin{equation}
    u(t) = K \exp\!\Big[ \ln\!\left(\dfrac{u_0}{K}\right)e^{-rt} \Big]. \label{eq:Gompertz_sol}
\end{equation}

\paragraph{The Richards' model.}
The Richards’ model~\cite{wang2012richards} generalises both the {l}ogistic and Gompertz models using a non-linear crowding function:
\begin{equation}
    g(u) = r \left(1 - \left(\dfrac{u}{K}\right)^{\nu}\right), \label{eq:Richards_g}
\end{equation}
where $r>0$ is the intrinsic growth rate, $K>0$ is the carrying capacity of the population, and $\nu>0$ is a shape parameter that modulates the strength of density dependence, controlling both the location of the inflection point and the degree of asymmetry in the growth dynamics.
The explicit solution, subject to the initial condition $u(0)=u_0$, is then given by
\begin{equation}
    u(t) = K \left[ 1 + \left(\left(\dfrac{K}{u_0}\right)^{\nu} - 1\right)e^{-\nu r t} \right]^{-1/\nu}. \label{eq:Richards_sol}
\end{equation}
If we consider the case where $\nu=1$, we recover the {l}ogistic growth law (Equation~\eqref{eq:log_g}) from Equation~\eqref{eq:Richards_g}.
Whereas as $\nu \to 0^+$ we find that the Richards’ model converges to the Gompertz growth law (Equation~\eqref{eq:Gompertz_g}). 
These rules are established analytically in \citep{tsoularis2002analysis}. 



\subsection{Data generation} \label{sec:data_gen}

To evaluate the proposed framework under controlled conditions, synthetic datasets were generated from the logistic, Gompertz, and Richards' models described above. 
In each case, the explicit solutions to Equation~\eqref{equation:PINNs} were used to {compute noise-free trajectories $u(t)$ from an initial condition $u(0)=u_0>0$.}
{The noise-free trajectories were evaluated at {twenty six} uniformly spaced time points over the interval $t \in [0,100]$. }
To mimic experimental variability, noise was subsequently added according to the power-law model in Equation~\eqref{eq:power_noise} and ten replicate datasets were generated for each configuration of parameters $(\sigma_0, \alpha)$.

\edit{For illustration, we consider three representative noise regimes in Figures~\ref{fig:intro}--\ref{fig:calib} spanning different forms and magnitudes of observation noise: additive \((\alpha=0,\sigma_0=0.1)\) for the logistic growth law, intermediate \((\alpha=0.5,\sigma_0=0.05)\) for the Gompertz growth law, and multiplicative \((\alpha=1,\sigma_0=0.2)\) for the Richards' growth law.
These parameter combinations are not intended to produce equivalent absolute noise magnitudes; rather, they were selected to demonstrate the ability of the NLL--BINN framework to recover dynamics across qualitatively different noise structures and severities. 
Alternatively, in Figure~\ref{fig:vs_rmse}, where direct comparisons between growth models are made, identical noise parameters are used across models. 
In Figure~\ref{fig:vs_rmse}, we compare results from all three models when the data is generated with multiplicative noise described by $\alpha=1$ and $\sigma_0=0.2$.}
{Unless otherwise stated, observations are clipped to ensure non-negative densities.}
The following parameter values are used throughout:
for the logistic model (Equation~\eqref{eq:log_sol}), $r=0.08$, $K=10$, $u_0=0.1$;
for the Gompertz model (Equation~\eqref{eq:Gompertz_sol}), $r=0.05$, $K=10$, $u_0=0.1$;
and for the Richards' model (Equation~\eqref{eq:Richards_sol}), $r=0.06$, $K=10$, $u_0=0.1$, and $\nu=2$.
{These parameter values were chosen such that all three models produce qualitatively similar sigmoidal growth dynamics over the observed time interval, whilst retaining distinctly different underlying crowding mechanisms. 
In particular, the values of $r$, $K$, and $u_0$ are selected to ensure comparable time scales and population ranges across models, enabling meaningful comparison of the learned dynamics. 
At the same time, differences in the functional form of the crowding term, and the inclusion of the additional shape parameter, $\nu$, in the Richards' model, lead to subtle but important variations in growth behaviour, including differences in asymmetry and the location of the inflection point. 
This makes the models sufficiently similar to present a challenging identification problem, whilst still allowing differences in mechanistic structure to be resolved.}
{The resulting simulated datasets are used consistently across all experiments presented in Sections~\ref{sec:recover}-\ref{sec:cal}.}

\section{Results}
In what follows, we evaluate the ability of the proposed NLL--BINN framework to jointly recover population dynamics, governing growth laws, and observation noise from noisy time series data. 
Using synthetic datasets generated from known models (see Section~\ref{sec:models}), {for which the ground truth is known,} we address three main questions:
\begin{enumerate}[(i)]
    \item can the framework accurately reconstruct the underlying dynamics and mechanistic growth functions;
    \item can it correctly identify and quantify different forms of observation noise; and
    \item does explicitly modelling the observation noise improve mechanistic recovery and uncertainty calibration?
\end{enumerate}

\subsection{Recovering mechanistic insights from noisy observations}\label{sec:recover}
We first assess whether the NLL--BINN framework can accurately recover both the underlying population dynamics and the associated crowding function from noisy observations {generated as described in Section~\ref{sec:data_gen}}.
Across all three growth models ({l}ogistic, Gompertz, and Richards’) described in Section~\ref{sec:models}, Figure~\ref{fig:intro} clearly shows that the NLL--BINN successfully reconstructs the smooth underlying solution $u(t)$ from noisy data (Figure~\ref{fig:intro}, top row), with the ensemble mean closely matching the ground truth. 
Importantly, the framework also recovers the corresponding crowding function $g(u)$ (Figure~\ref{fig:intro}, middle row \edit{and Supplementary Figure~\ref{fig:rlow} for low-noise Richards' data}), demonstrating that the framework is capable of identifying the governing dynamical law without prescribing the functional form \textit{a priori}.

\begin{figure}[htbp]
    \centering
    \includegraphics{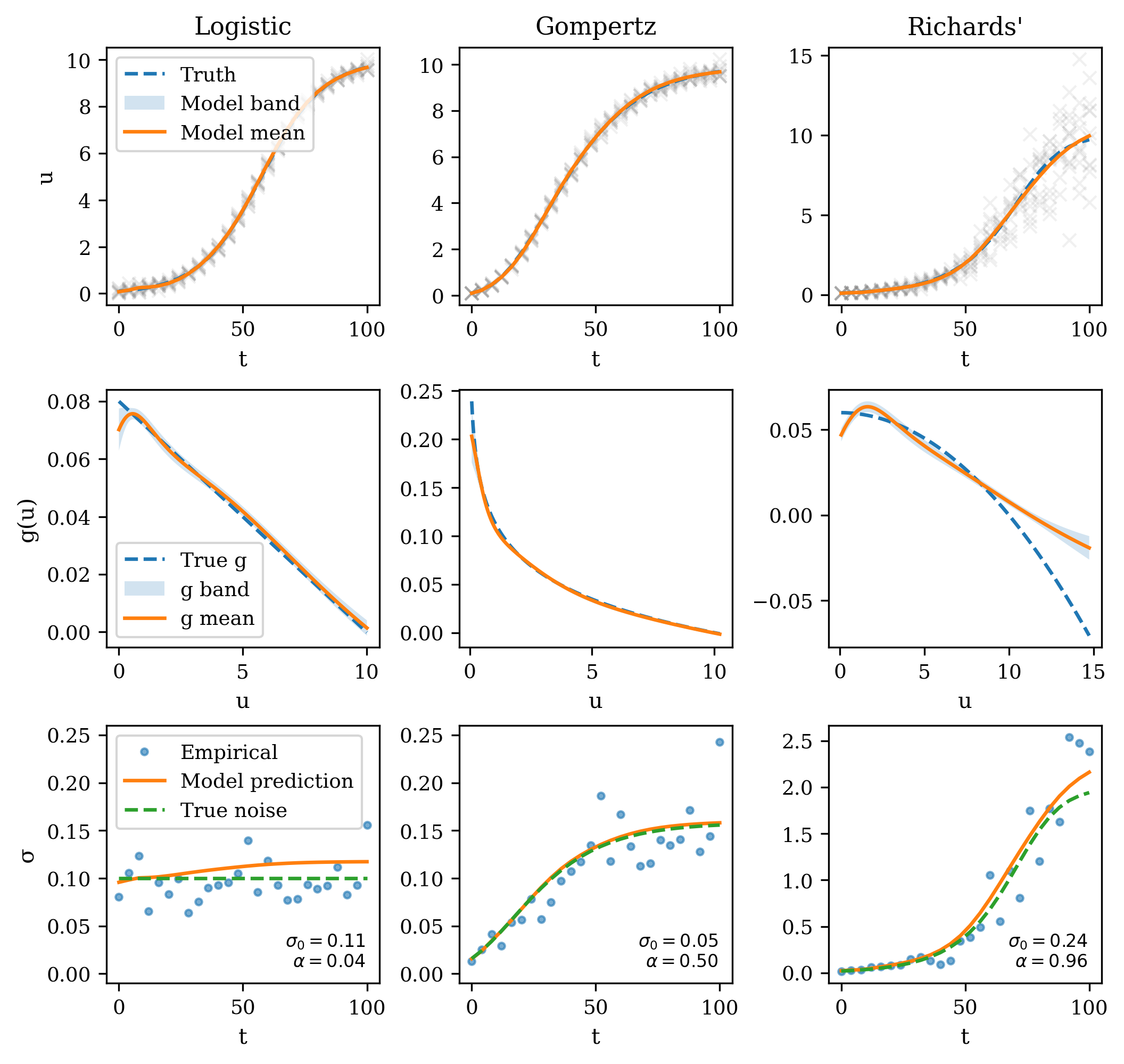}
    \caption{Plots demonstrating the recovery of population dynamics, growth laws, and noise structure using the NLL--BINN framework for {synthetic data generated from the} {l}ogistic, Gompertz, and Richards’ models {(Section~\ref{sec:models}) with parameters $r$, $K$, $u_0$, and $\nu$ as specified in Section~\ref{sec:data_gen}. Noise was added according to Equation~\eqref{eq:data_struc}, using $\alpha = 0$ and $\sigma_0=0.1$ for the Logistic growth model, $\alpha = 0.5$ and $\sigma_0=0.05$ for the Gompertz model, and $\alpha = 1$ with $\sigma_0=0.2$ for the Richards' model.} 
    {For clarity, the logistic, Gompertz, and Richards datasets shown in this figure also correspond to the additive, intermediate, and multiplicative noise models shown in Figures~\ref{fig:params_noise} and~\ref{fig:calib}, respectively.}
    Top row: mean learned trajectory $u_\theta(t)$ in solid orange lines, with one standard deviation (light blue) and noisy observations from all ten replicates (grey). Middle row: inferred crowding functions $g_\phi(u)$ with the mean shown in orange, the truth as a dashed blue line and the variability shown by blue shading. Bottom row: learned noise profiles, with the empirical results as blue dots\edit{, the true noise model in green dashed lines} and the average learned noise model across all repetitions in orange.
    }
    \label{fig:intro}
\end{figure}

It can also be observed in Figure~\ref{fig:intro} that there are small discrepancies observed in the learned growth laws in regions of the domain with small, almost zero, population densities. 
In particular, this uncertainty and corresponding wide-ranging predictions are, perhaps, most clear for the {l}ogistic growth law (Figure~\ref{fig:intro}, middle row, left).  
By looking at the structure of the true underlying ODE that the NLL--BINN is trying to learn, it is clear why there is higher variability in these regions: 
Equation~\eqref{equation:PINNs} ensures $u(t)\geq u_0>0$ for all $t\geq0$ as long as $g(u; \boldsymbol{p})\geq0.$
However, the BINN attempts to learn the growth laws for all non-negative densities.
As such, for all values $0\leq u<u_0$, there are no data points within the {observations} that the BINN can use during optimisation.
Instead the training in this region is purely being driven by the ODE and biological loss terms, for which there are many possible solutions, including those shown in Figure~\ref{fig:intro}.

In addition, higher variability can also be observed for the Richards' model for both large and small $u$ values.
{At low densities, this {variability} {again} reflects a lack of {observations in the} training data, as discussed above. 
At high densities, however, {the variability} arises for a different reason: in the noise-free system, the solution is bounded above by the carrying capacity, $K$, and thus no true data exists for $u > K$. 
However, the presence of training data in this region is solely a consequence of the added noise, which perturbs measurements beyond the biologically meaningful range. 
This effect is most pronounced for the Richards dataset, which also {has} the largest noise magnitude ($\sigma_0 = 0.2$), leading to greater dispersion of observations outside the physically relevant domain.
As a result, the NLL--BINN is exposed to data in regions of the state space that are not supported by the underlying dynamics, leading to inferred dynamics there are not strongly constrained by the true system, increasing variability and reducing identifiability.}
\edit{An example of better constrained learning to demonstrate this principle using data generated using the Richards' growth law with lower noise levels can be found in Supplementary Figure~\ref{fig:rlow}.}



In the bottom row of Figure~\ref{fig:intro}, we also see that in addition to recovering the underlying solution and growth dynamics, the NLL--BINN is also able to simultaneously learn the noise model, $\sigma(u)$, from the data. 
Taken together, these results demonstrate that the NLL--BINN framework can successfully disentangle underlying trajectories, governing dynamics, and noise structure simultaneously whilst being trained on noisy time-series data.

\subsection{Identification of heteroscedastic noise models}

We next evaluate whether the model can correctly identify different forms of heteroscedastic noise {using the synthetic datasets described in Section~\ref{sec:data_gen}.}
Figure~\ref{fig:params_noise} (left) shows the learned variance for each dataset as a function of the population density. 
In each case, the inferred variance closely follows the true {underlying relationship in the inputted data}, with almost perfect constant variance observed for additive noise, and density-dependent scaling noticeable for both the multiplicative and intermediate cases.

These results are further supported by the noise profiles shown in the bottom row of Figure~\ref{fig:intro}. 
In the {l}ogistic example, the inferred noise {remains approximately constant across the trajectory}.
In the Richards' example {(multiplicative noise)}, {the noise magnitude increases with population density, matching the empirical spread in the data.}
{The Gompertz model (intermediate noise) exhibits behaviour between these two extremes, demonstrating that the framework can successfully interpolate between noise regimes.}

\begin{figure}[h!]
    \centering
    \includegraphics{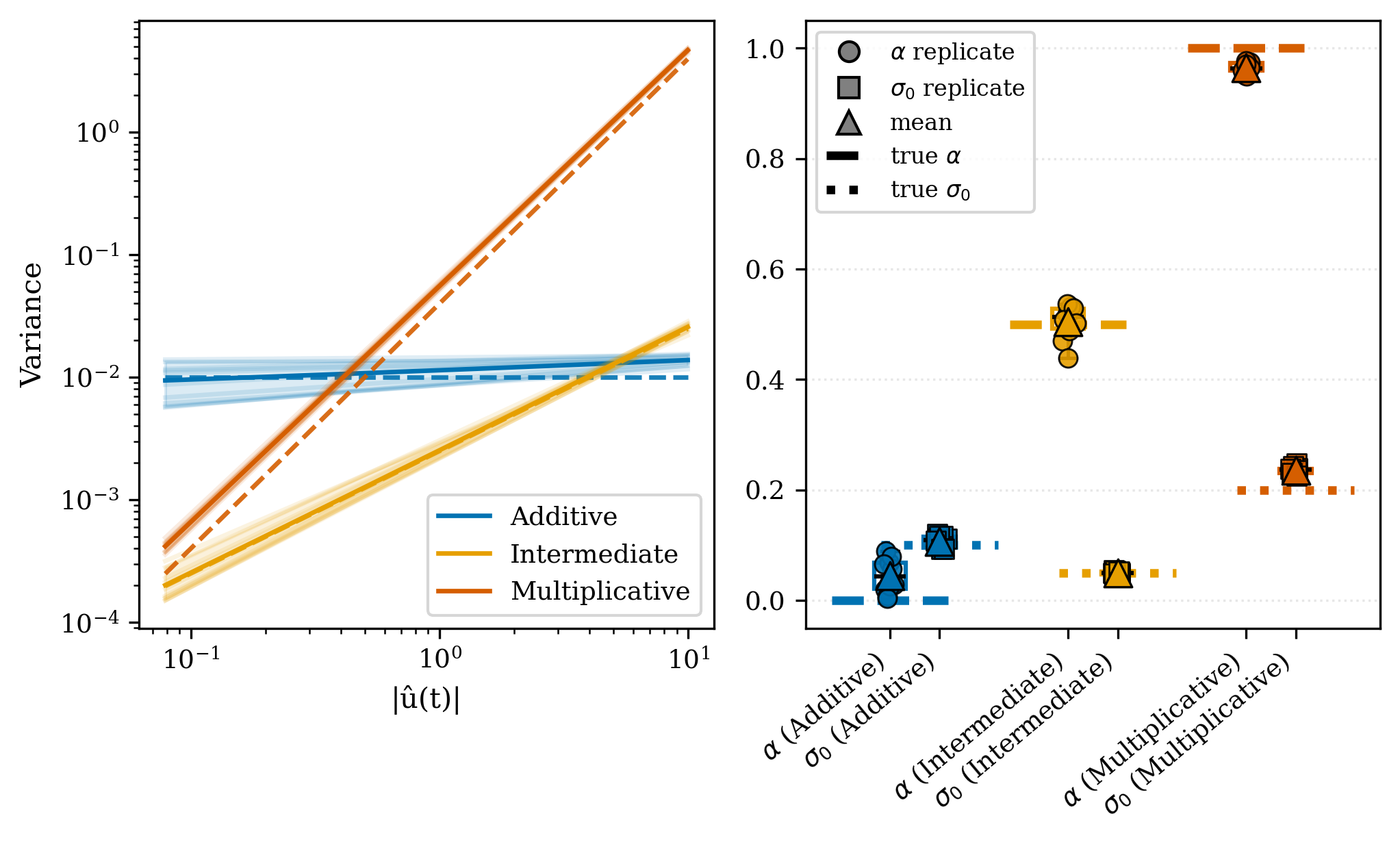}
    \caption{Left: learned variance $\sigma^2(u)$ as a function of the population density, $u$, for ten replicated synthetic datasets generated from the logistic, Gompertz, and Richards' models (Section~\ref{sec:models}) with parameters $r$, $K$, $u_0$, and $\nu$ as specified in Section~\ref{sec:data_gen}. 
    \edit{Noise was added according to Equation~\eqref{eq:data_struc}, using $\alpha = 0$ and $\sigma_0=0.1$ for the logistic growth model (additive, \edit{blue}), $\alpha = 0.5$ and $\sigma_0=0.05$ for the Gompertz model (intermediate, \edit{yellow}), and $\alpha = 1$ with $\sigma_0=0.2$ for the Richards' model (multiplicative, \edit{orange}).} 
    {For clarity, the logistic, Gompertz, and Richards datasets shown in Figure~\ref{fig:intro} correspond{,} respectively{,} to the additive, intermediate, and multiplicative noise models shown in this figure and Figure~\ref{fig:calib}.}
    Right: inferred noise parameters ($\sigma_0,\, \alpha$) across ten ensemble runs plotted against their true values (additive in \edit{blue}, multiplicative in \edit{yellow} and intermediate in \edit{orange}). True $\alpha$ values are plotted as dashed lines, whereas true $\sigma_0$ values are plotted as dotted lines.}
    \label{fig:params_noise}
\end{figure}

Quantitatively, the learned parameters ($\sigma_0,\, \alpha$) also closely match their ground truth values across all noise regimes (Figure~\ref{fig:params_noise}, right), with limited variability observed between ensemble members. 
In particular, for additive noise, $\alpha \approx 0$, indicating correct identification of homoscedastic variance.
For multiplicative noise, $\alpha \approx 1$, demonstrating recovery of linear variance scaling with the population density{, whilst NLL--BINNs trained on data with intermediate heteroscedastic noise {also} accurately recover}
the correct parameter values, showing that the model can interpolate between noise regimes.

A small bias is observed in the additive noise case, where the learned $\alpha$ deviates slightly from zero.
This can be attributed to the ``clipping" of the data at zero, which prevents negative (and biologically unrealistic) densities and subsequently introduces a form of censoring that distorts the underlying Gaussian noise distribution, particularly at low densities.
As such, the negative contributions from the full additive Gaussian noise are suppressed, leading to an apparent reduction and asymmetry in the noise near the boundary.
Within the power-law parameterisation, this distortion can be absorbed by small deviations in $\alpha$, especially close to $\alpha = 0$ where changes in $\alpha$ have only a weak effect on the variance.
When negative values are retained (\textit{i.e.}, in the absence of clipping), this bias is substantially reduced, as shown in Supplementary Figure~\ref{fig:clipping}, confirming that these small errors arise only as a consequence of the imposed data structure, rather than the NLL--BINN framework itself.
Despite this, the inferred variance remains effectively constant, and the overall noise structure is still accurately captured, demonstrating that the NLL--BINN can reliably distinguish and quantify different noise mechanisms and amplifications directly from data.

\subsection{Uncertainty calibration from likelihood-based training}\label{sec:cal}
Beyond recovering the functional form of the noise and the underlying growth dynamics, it is important to assess whether the learned uncertainty is well calibrated.
{In a well-calibrated model with approximately Gaussian predictive uncertainty, around 68\% of observations should lie within one predicted standard deviation, and around 95\% within two standard deviations{,} of the mean.}

\begin{figure}[h!]
    \centering
    \includegraphics{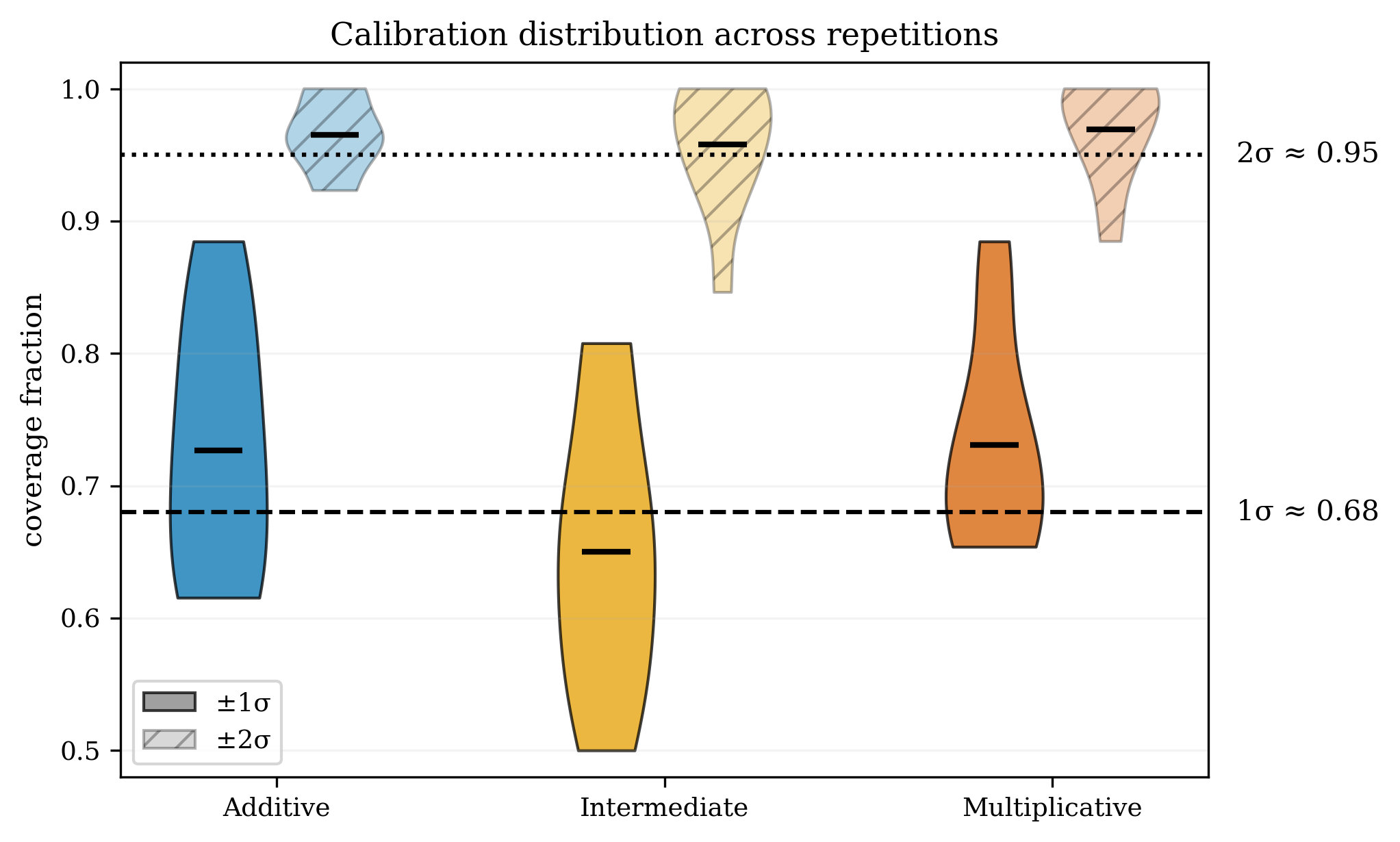}
    \caption{Plot showing the calibration of predictive uncertainty for ten replicated synthetic datasets generated from the logistic, Gompertz, and Richards' models (Section~\ref{sec:models}) with parameters $r$, $K$, $u_0$, and $\nu$ as specified in Section~\ref{sec:data_gen}. 
    \edit{Noise was added according to Equation~\eqref{eq:data_struc}, using $\alpha = 0$ and $\sigma_0=0.1$ for the logistic growth model (additive, \edit{blue}), $\alpha = 0.5$ and $\sigma_0=0.05$ for the Gompertz model (intermediate, \edit{yellow}), and $\alpha = 1$ with $\sigma_0=0.2$ for the Richards' model (multiplicative, \edit{orange}).} 
    {For clarity, the logistic, Gompertz, and Richards datasets shown in Figure~\ref{fig:intro} correspond{,} respectively{,} to the additive, intermediate, and multiplicative noise models shown in this figure and Figure~\ref{fig:calib}.}
    The proportion of observations contained within predicted confidence intervals is shown for multiple coverage levels. Dashed lines indicate the ideal calibration and hold across all noise models. 
    }
    \label{fig:calib}
\end{figure}

Figure~\ref{fig:calib} shows the empirical coverage of the predicted uncertainty intervals for ten repetitions. 
Across all noise regimes, the observed coverage closely matches these theoretical values, indicating that the learned uncertainty is well calibrated. 
These calibration results holds across additive, intermediate, and multiplicative noise models, demonstrating that the framework generalises across different heteroscedastic noise structures.
Importantly, calibration is achieved without explicitly enforcing coverage constraints during training.
Instead, these arise naturally from the likelihood-based formulation of the data loss function in Equation~\eqref{eq:nll_loss}. 
Together, these results show that the model provides both accurate predictions and reliable uncertainty quantification.

\subsection{The impact of likelihood-based training on mechanistic model recovery}
Finally, to examine the impact of explicitly modelling and learning the observation noise present within the data, we compare NLL--BINNs to the more commonplace BINN that is trained using a standard root mean squared error (RMSE)-based loss term. 
We assess the {model performance by computing} the RMSE between the true underling dynamics and forward simulations using the learned growth law, which we call the \emph{mechanistic error}. 

{This metric differs from a standard measure of data misfit. Rather than measuring the agreement between the simulated solution and the noisy observations, the mechanistic error evaluates how accurately the inferred governing equations reproduce the true noise-free dynamics.
This distinction is important in our setting, where the goal is not merely to interpolate observed data, but to recover the true underlying biological mechanism.
In particular, models trained with an RMSE loss may overfit observation noise, leading to accurate data fits but biased or unstable dynamics.
By contrast, the mechanistic error provides a direct measure of how well the learned model generalises beyond the observed data and captures the true system behaviour.}

\begin{figure}[h!]
    \centering
    \includegraphics{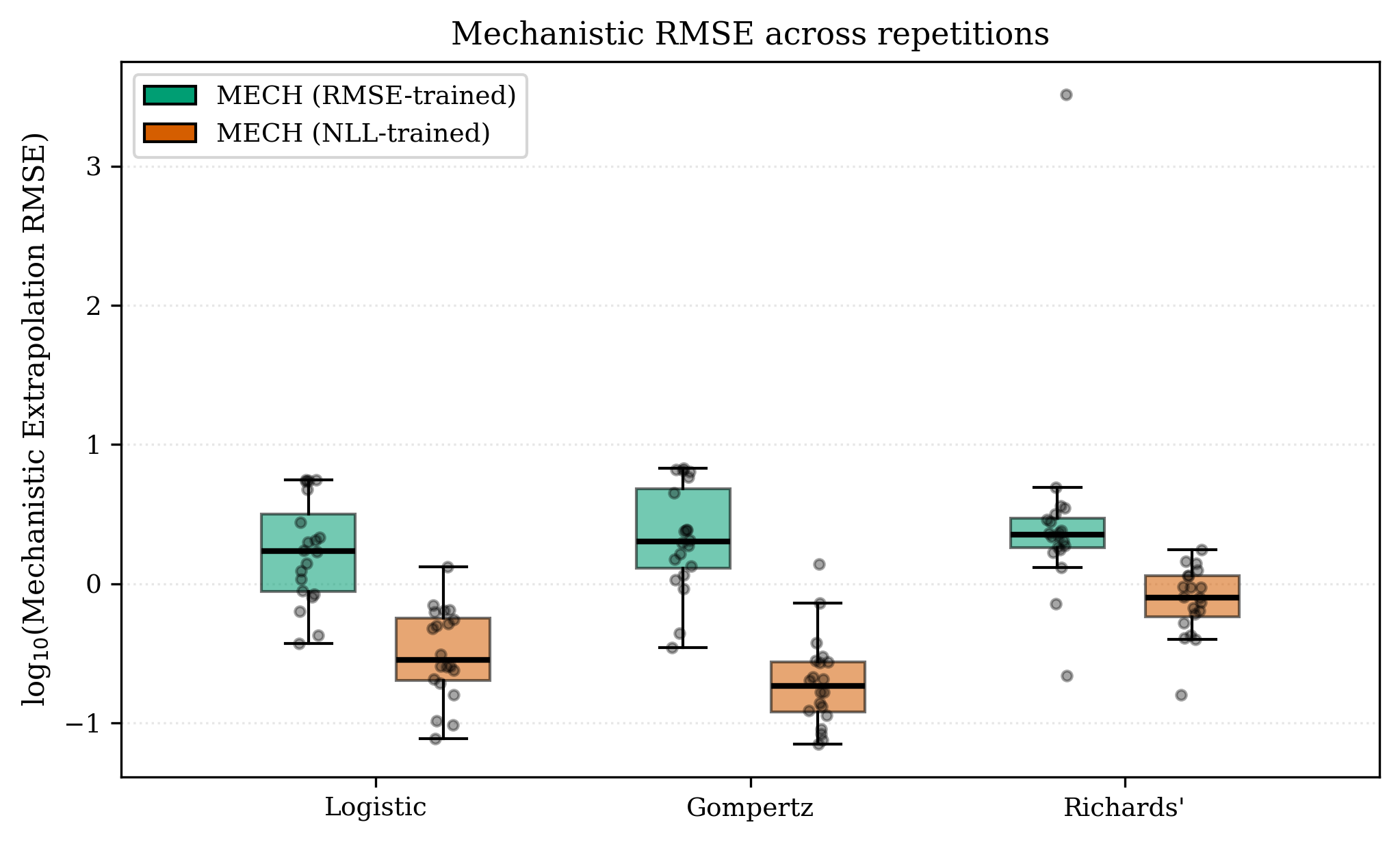}
    \caption{
    {Plot} showing the distribution of the root mean squared error between the true underlying dynamics and the mechanistically simulated solutions using the learned growth dynamics across ten {replicated synthetic datasets generated from the logistic, Gompertz, and Richards' models (Section~\ref{sec:models}) with parameters $r$, $K$, $u_0$, and $\nu$ as specified in Section~\ref{sec:data_gen}. Noise was added \edit{to all datasets} according to Equation~\eqref{eq:data_struc}, using $\alpha = 1$ and $\sigma_0=0.2$ \edit{to simulate multiplicative noise in all cases}. } 
    Results are tested for both the NLL--BINN introduced in this work (orange) and a standard BINN, trained using a root mean squared error (RMSE, green) for the data loss. 
    A seed is set for each repetition such that the data split into training, test and validation subsets is the same for both BINNs, as well as the initialisation. 
    Thereafter, differences in the training process and overall results are driven purely by the differences in the data loss functions for the BINNs (NLL, or RMSE).
    }
    \label{fig:vs_rmse}
\end{figure}

Figure~\ref{fig:vs_rmse} shows the distribution of the mean mechanistic error across ten simulated datasets for each growth model.
{For each dataset, \edit{twenty} NLL--BINN models were independently trained, after which the learned dynamics were forward-simulated and compared to the ground-truth solution to compute the mechanistic error.
This post-training evaluation isolates how well the inferred governing equations reproduce the true dynamics, rather than how well the models fit noisy observations.}

{While improvements are pronounced for the logistic and Gompertz models, gains for the Richards' model are more modest.
This is expected due to its increased structural flexibility: the additional shape parameter $\nu$ introduces a continuum of growth behaviours, which can produce similar trajectories for different parameter combinations.
Such practical non-identifiability makes it more difficult to uniquely recover the underlying dynamics from noisy data, leading to higher variability in the learned models and reduced improvements in mechanistic accuracy \cite{simpson2022parameter}.}

\edit{Furthermore, to investigate the role of specifically learning a flexible noise model with parameters that can adapt during neural network training, we compared the results of our NLL--BINN approach to a two-stage approach using the trajectories learned from a standard RMSE--based BINN, followed by linear regression of the residuals to a power law noise model. 
Specifically, for each ensemble member, we regressed $\log |y-\hat{u}|$ against $\log|\hat{u}|$, corresponding to the power-law noise model in Equation~\eqref{eq:power_noise}.
In Figure~\ref{fig:new}, we see that the two-stage approach does recover a clear state-dependent residual structure across all three growth laws considered, demonstrating that a post-hoc approach can indeed identify some features of the noise. However, it systematically deviates from the true noise law. In contrast, the jointly trained NLL-BINN more closely recovers the prescribed noise structure across the full range of $u$. }

\edit{These results suggest that whilst the residuals of an independently fitted BINN contain information about the observation noise, post-hoc residual fitting does not fully separate observational variability from errors in the reconstructed trajectory. 
Joint NLL--BINN training instead estimates the solution and noise model simultaneously, allowing the likelihood contribution of the observations to influence the inferred trajectory and noise structure together.}

\begin{figure}[htbp]
    \centering
    \includegraphics[width=\linewidth]{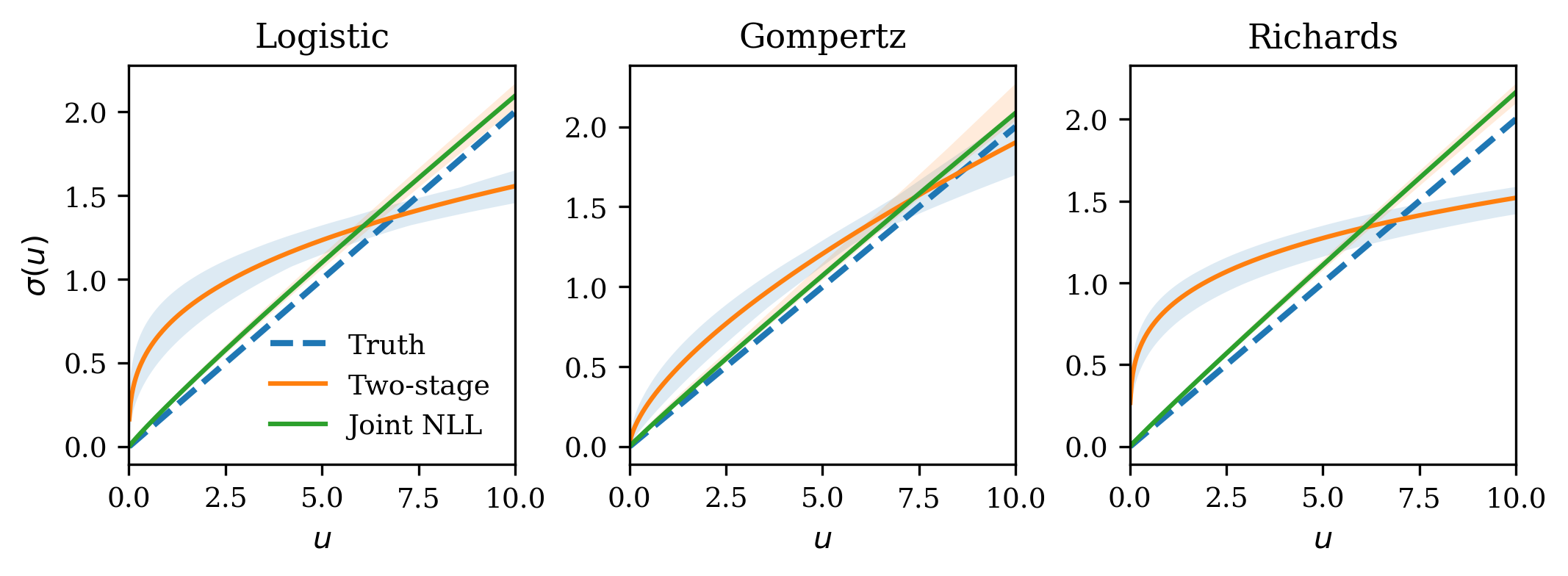}
    \caption{\edit{Plots of the noise standard deviation, $\sigma(u)$, inferred using the NLL--BINN (green, with shaded orange 95\% intervals) and using a two-stage approach where we learn with a RMSE--BINN and fit a power-law model (Equation~\eqref{eq:power_noise} after using linear regression of residuals residuals, $\log|y-{u}|$ against $\log|{u}|$ (orange, with shaded blue 95\% intervals), against the true noise relationship used to generate the synthetic data (blue dashed) for the Logistic, Gompertz, and Richards models (Section~\ref{sec:models}).}}
    \label{fig:new}
\end{figure}

{Overall, w}e find that the NLL--BINN consistently achieves a lower mechanistic error compared to its RMSE--trained counterpart. 
As a result, both the median error and variability between experimental repetitions are reduced, indicating that incorporating a probabilistic, learnable noise model improves both the accuracy and robustness of the inferred dynamics. 
This improvement is consistent with the interpretation of the NLL-BINN as a correctly specified penalised likelihood approach, which weights observations according to their noise level, in contrast to RMSE-based training that \edit{does not optimise for the noise model parameters simultaneously during neural network training.}
In particular, the reduction in variability suggests that likelihood-based training stabilises the learning process, leading to more reproducible recovery of the governing equations.




{\subsection{An example: coral reef data from Lady Musgrave Island}}\label{sec:CR}
{To demonstrate the applicability of the NLL--BINN framework to real-world data, we consider measurements of hard coral cover at two reef sites near Lady Musgrave Island, Australia, where regrowth was observed following disturbance~\cite{simpson2022parameter}. 
These data exhibit characteristic sigmoidal re-growth dynamics, with an initial phase of rapid increase followed by a gradual approach to a saturation level.}
{In contrast to the synthetic datasets considered previously, these observations consist of single trajectories without replicated measurements, meaning that the variability in the data cannot be estimated independently. 
As a result, both the underlying dynamics and the observation noise must be inferred simultaneously from limited data, presenting a challenging setting for mechanistic learning.}

{Figure~\ref{fig:coral} shows the results of applying the NLL--BINN framework to {data from} both sites. 
In each case, the learned trajectory, $u_\theta(t)$ (left column), provides a smooth representation of the observed data, capturing the expected sigmoidal regrowth. 
The inferred dynamics are further characterised through the learned crowding function, $g_\phi(u)$ (middle column), which exhibits a decreasing relationship with population density, consistent with density-limited growth. 
Notably, the shape of $g_\phi(u)$ differs between the two sites, suggesting variations in the underlying growth mechanisms or environmental conditions.}

\begin{figure}[htbp]
    \centering
    \includegraphics[width=\linewidth]{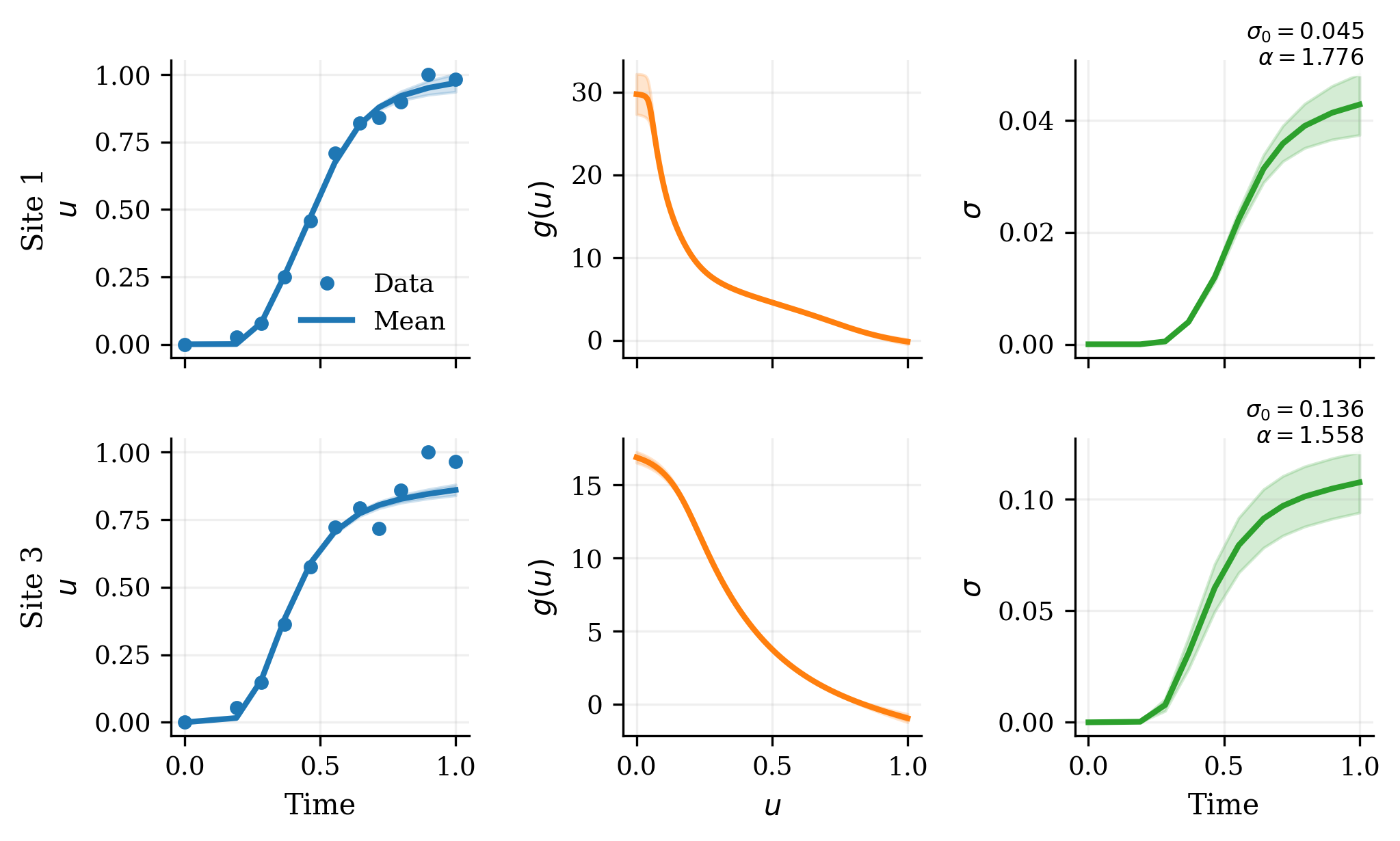}
    \caption{Application of the NLL--BINN framework to coral re-growth data from two reef sites near Lady Musgrave Island, Australia (top: Site 1; bottom: Site 3). Left column: learned trajectories $u_\theta(t)$ (solid blue) fitted to the observed data (points), showing sigmoid growth curves. 
    Middle column: inferred crowding functions $g_\phi(u)$, capturing the density-dependent growth behaviour without prescribing a functional form. 
    Right column: learned state-dependent noise models $\sigma(u)$, with shaded regions indicating variability across the ensemble. In both sites, the inferred noise increases with population density, consistent with heteroscedastic variability. 
    Estimated noise parameters $(\sigma_0, \alpha)$ are shown for each site.}
    \label{fig:coral}
\end{figure}

{In addition to recovering the dynamics, the framework simultaneously infers a density-dependent noise model, $\sigma(u)$ (right column). 
For both sites, the learned noise increases with population density, indicating heteroscedastic variability in the observations. 
This behaviour is consistent with a multiplicative-like noise structure, although the inferred exponents ($\alpha \approx 1.56$ and $\alpha \approx 1.78$) suggest stronger-than-linear scaling in these cases. 
Crucially, this information is extracted directly from the data without requiring replicated experiments, providing a data-driven estimate of observational uncertainty \edit{under the assumed power law observation model}.}
\edit{Notably, since the dataset does not contain replicate observations, the data provide limited information for distinguishing among competing variance functions. In general cases, the choice of the most appropriate noise model, and its fit to data, may be assessed using residual diagnostics, held-out likelihood, information criteria, and predictive-interval calibration.}


{Overall, th\edit{e coral reef data} example demonstrates that the NLL--BINN framework can be effectively applied to real experimental datasets, even in settings where both the noise structure and governing dynamics are unknown. 
By jointly learning these components within a likelihood-based framework, the method provides a flexible and interpretable approach for analysing biological growth processes in the presence of heteroscedastic and unobserved variability, while also highlighting the fundamental role of data availability in determining model identifiability.
In particular, the single-trajectory nature of the coral reef data limits the ability to uniquely identify the underlying growth mechanisms, meaning that the inferred dynamics should be interpreted as one of several plausible explanations consistent with the observed data.}

\section{Discussion}


In this work, we introduced an extension to the BINN framework that enables the joint learning of system dynamics and observation noise structure directly from data. 
By incorporating a density-dependent noise model within a likelihood-based training {process}, the proposed approach moves beyond standard neural ODE formulations that assume fixed, homoscedastic noise \edit{in advance to allow for simultaneous learning of noise-related parameters during neural network training.}
Across a range of synthetic, sigmoidal population growth models, we demonstrated that the framework can accurately recover both the underlying mechanistic laws and the functional form of heteroscedastic variability.

A key conceptual contribution of this work is the treatment of noise as a mechanistic quantity rather than a nuisance. 
In many biological systems, variability reflects intrinsic heterogeneity, stochastic effects, or measurement processes that scale with population size. 
By learning a functional relationship between {the} noise magnitude and population density, the {approach} provides insight into how uncertainty arises from the underlying system or measurement process. 
In this sense, the inferred noise model complements the learned dynamics, offering an additional layer of interpretability that is typically absent in standard neural ODE approaches.

The results highlight several practical advantages. 
Firstly, learning a density-dependent noise model leads to improved uncertainty quantification, as the model captures heteroscedastic variability and produces well calibrated predictive intervals. 
Secondly, the likelihood-based formulation naturally weights observations according to their noise level, improving robustness to noisy or unevenly distributed data.
Thirdly, the use of an interpretable noise parametrisation allows the model to distinguish between additive, multiplicative, and intermediate noise regimes, providing a direct link between statistical variability and mechanistic interpretation.

%
{One} key consideration in jointly learning dynamics and noise is the question of identifiability. 
In principle, there exists the possibility of trade-offs between the learned dynamics and the noise model.
For example, in Figure~\eqref{fig:intro} we observed that regions with sparse data, such as very low or high population densities, lead to increased uncertainty and reduced identifiability. 
So far we have mitigated such effects through the combined use of ODE constraints, biologically motivated regularisation, and ensemble training, which together restrict the space of admissible solutions. 
Empirically, we observe consistent recovery of both dynamics and noise structure across repeated experiments, suggesting that the formulation is sufficiently well-posed under the data regimes considered here.
{However, this also highlights an important practical consideration: accurate recovery of mechanistic model structure depends not only on the quantity of data, but also on its distribution within the physically relevant domain. 
In particular, care should be taken when interpreting learned dynamics in regions that are only populated due to noise, as these may not reflect the true behaviour of the system.
As such}, a full characterisation of identifiability in this setting---particularly its dependence on data sparsity, noise magnitude, and model parametrisation---remains an open question. 

Other natural extensions to this work exist. 
{While} the power-law noise model provides a flexible and interpretable parametrisation, it may not capture more complex noise structures present in real systems, such as {the} one-sided noise observed in this work for small densities.
{More generally, any parametric noise distribution for which the negative log-likelihood can be evaluated could be incorporated within this framework{ using its own flexible function approximating neural network}.
Bayesian extensions of the BINN framework could then further improve uncertainty quantification by accounting for parameter uncertainty in both the dynamics and noise models. 

{More broadly, the proposed framework is compatible with a wide class of neural differential equation approaches, including universal differential equations, where neural networks are embedded within partially specified mechanistic models~\cite{loman2025functional, rackauckas2020universal}. 
In such settings, the ability to learn density-dependent noise alongside unknown dynamical components may provide a principled route to improving both parameter inference and uncertainty quantification, particularly in systems where variability is intrinsic to the underlying process.}
\edit{Whilst also, this approach is amenable for applications to multivariable systems of equations, or in the context of partial differential equations (PDEs) where limited adaptation of the underlying framework would be required: either additional inputs (in the context of PDEs) or replicate MLPs with their own parameters (in the context of multi-species sytems).}

Overall, this work demonstrates that incorporating learnable, density-dependent noise into {mechanistic neural differential equation frameworks} provides a {simple, yet powerful, approach} for modelling heteroscedastic uncertainty in dynamical systems. 
By jointly learning {the underlying} dynamics and {data} variability, the approach {presented in this work} advances interpretable machine learning methods for noisy biological data, and offers a principled foundation for integrating uncertainty into data-driven mechanistic inference.

\paragraph{Acknowledgements.}
R.M.C. acknowledges support from the Engineering and Physical Sciences Research Council (EP/Z534870/1). 
R.E.B. acknowledges support of a grant from the Simons Foundation (MP-SIP-00001828). 
For the purpose of Open Access, the authors have applied a CC BY public copyright licence to any Author Accepted Manuscript (AAM) version arising from this submission. 

\paragraph{Declarations.}
All data \edit{and code used to generate results} in this work is available online at the following GitHub link: \url{github.com/beckycrossley/Learning-noise-growth-BINNs}.
The authors declare that they have no competing interests. 

\bibliographystyle{unsrt}
\bibliography{refs}

\begin{center}\
    \textbf{\Large{Supplementary Figures}}
\end{center}

\setcounter{figure}{0}  
\renewcommand{\thefigure}{S\arabic{figure}}

\setcounter{table}{0}  
\renewcommand{\thetable}{S\arabic{table}}

\setcounter{equation}{0}  
\renewcommand{\theequation}{S\arabic{equation}}

\setcounter{section}{0}
\renewcommand{\thesection}{S\arabic{section}}
\begin{figure}[h!]
    \centering
    \includegraphics[width=\linewidth]{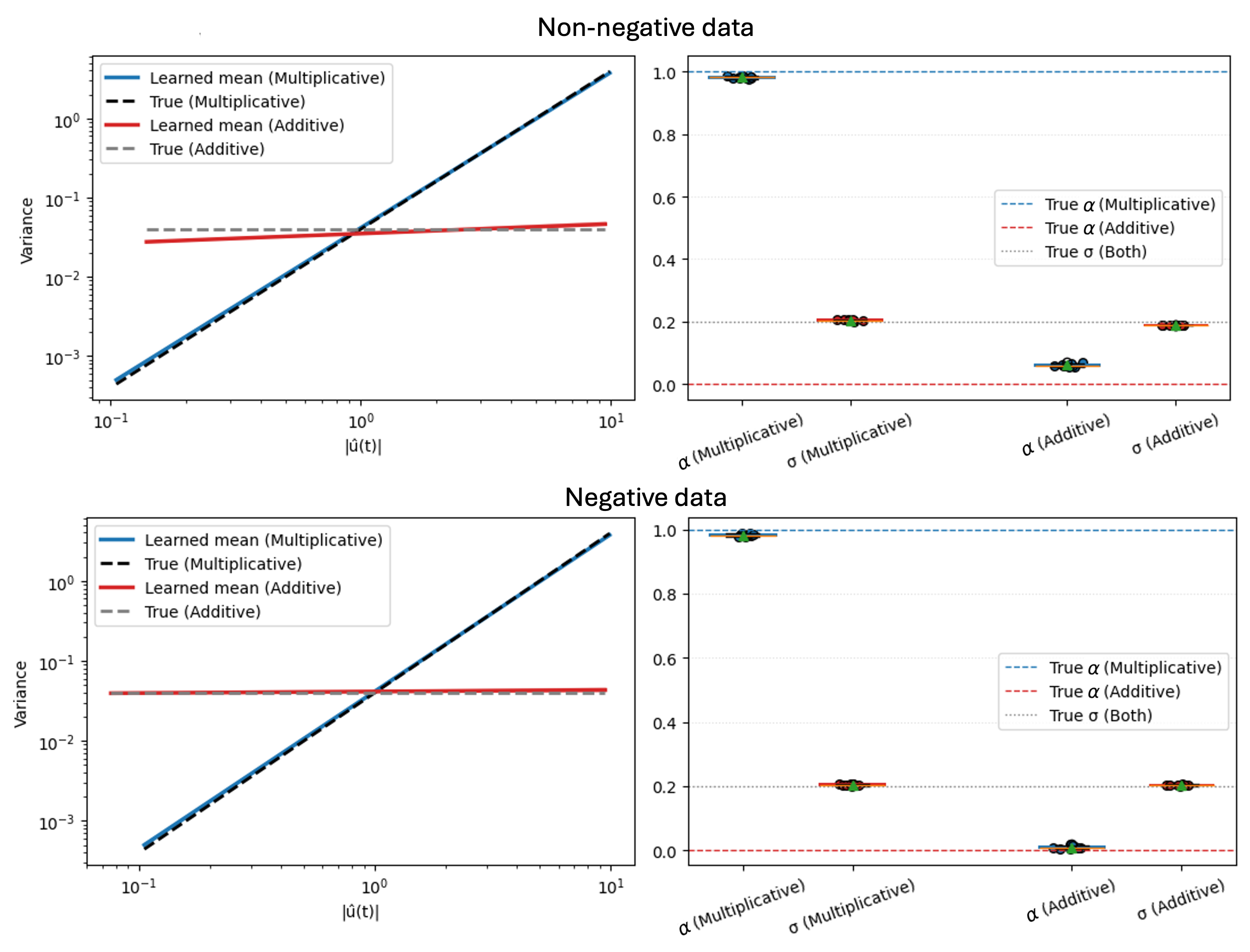}
    \caption{Left: learned variance $\sigma^2(u)$ as a function of the population density, $u$, with true underlying constant variance for additive noise ($\alpha = 0$, red) and state-dependent scaling for multiplicative noise ($\alpha = 1$, blue). 
    Right: inferred noise parameters ($\sigma_0,\, \alpha$) across ten ensemble runs plotted against their true values (additive in red, multiplicative in blue). 
    The top row shows example results for {one hundred and one time points of} input data that did not contain any negative values. The bottom row shows {the same} results when negative densities were allowed. These results demonstrate that the discrepancies observed in the results for additive data arise as a result of the data structure, rather than the BINN framework itself.}
    \label{fig:clipping}
\end{figure}
\begin{figure}
    \centering
    \includegraphics[width=\linewidth]{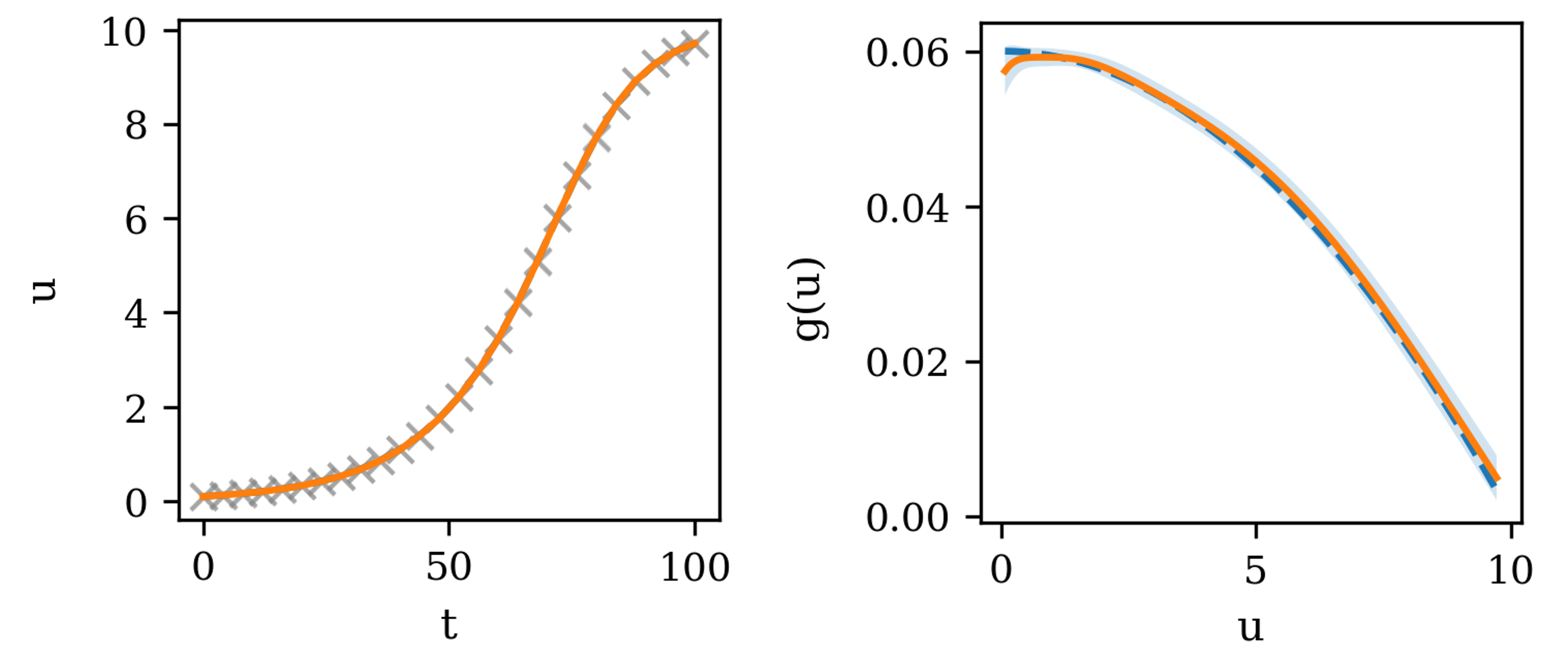}
    \caption{\edit{Plots demonstrating the recovery of the population dynamics and growth law using the NLL--BINN framework for {synthetic data generated from the} Richards’ model {(Section~3) with parameters $r$, $K$, $u_0$, and $\nu$ as specified in Section~3.1. Noise was added according to Equation~(3), using $\alpha = 0$ and $\sigma_0=0.01$.} 
    Left: mean learned trajectory $u_\theta(t)$ in solid orange lines, with one standard deviation (light blue) and noisy observations from all ten replicates (grey). Right: inferred crowding functions $g_\phi(u)$ with the mean shown in orange, the truth as a dashed blue line and the variability shown by blue shading.}}
    \label{fig:rlow}
\end{figure}
\end{document}